\documentclass[sigconf,nonacm]{acmart}
\usepackage{tikz}

\AtBeginDocument{%
  \providecommand\BibTeX{{%
    \normalfont B\kern-0.5em{\scshape i\kern-0.25em b}\kern-0.8em\TeX}}}

\makeatletter
\def\@ACM@checkaffil{
    \if@ACM@instpresent\else
    \ClassWarningNoLine{\@classname}{No institution present for an affiliation}%
    \fi
    \if@ACM@citypresent\else
    \ClassWarningNoLine{\@classname}{No city present for an affiliation}%
    \fi
    \if@ACM@countrypresent\else
        \ClassWarningNoLine{\@classname}{No country present for an affiliation}%
    \fi
}
\makeatother

\begin{document}

\title{S3C2 Summit 2022-09: \\ Industry Secure Supply Chain Summit}

\author{Mindy Tran$^{*}$, Yasemin Acar$^{*}$, Michel Cukier$^{\dagger}$, William Enck$^{\ddagger}$,\\ Alexandros Kapravelos$^{\ddagger}$, Christian Kästner$^{\mathsection}$, Laurie Williams$^{\ddagger}$}

\def \authors{Mindy Tran, Yasemin Acar, Michel Cukier, William Enck, Alexandros Kapravelos, Christian Kästner, Laurie Williams}

\affiliation{%
    \institution{$^*$Paderborn University, Paderborn, Germany, and George Washington University, DC, USA}
}
\affiliation{%
    \institution{$^\dagger$University of Maryland, College Park, MD, USA}
}
\affiliation{%
    \institution{ $^\ddagger$North Carolina State University, Raleigh, NC, USA}
}
\affiliation{%
    \institution{ $^\mathsection$Carnegie Mellon University, Pittsburgh, PA, USA}
}

\renewcommand{\shortauthors}{Secure Software Supply Chain Center (S3C2)}
\renewcommand{\shorttitle}{S3C2 Summit 2022-09: Industry Secure Supply Chain Summit}

\begin{abstract}
  Recent years have shown increased cyber attacks targeting less secure elements in the software supply chain and causing fatal damage to businesses and organizations. Past well-known examples of software supply chain attacks are the SolarWinds or log4j incidents that have affected thousands of customers and businesses. The US government and industry are equally interested in enhancing software supply chain security. We conducted six panel discussions with a diverse set of 19 practitioners from industry. We asked them open-ended questions regarding SBOMs, vulnerable dependencies, malicious commits, build and deploy, the Executive Order, and standards compliance. The goal of this summit was to enable open discussions, mutual sharing, and shedding light on common challenges that industry practitioners with practical experience face when securing their software supply chain. This paper summarizes the summit held 
  on September 30, 2022. 
  Full panel questions can be found at the beginning of each section and in the appendix.
\end{abstract}

\keywords{software supply chain, open source, secure software engineering}



\maketitle

\begin{tikzpicture}[overlay, remember picture]
\node[anchor=north west, 
      xshift=17.5cm, 
      yshift=-2.1cm] 
     at (current page.north west) 
     {\includegraphics[width=2.1cm]{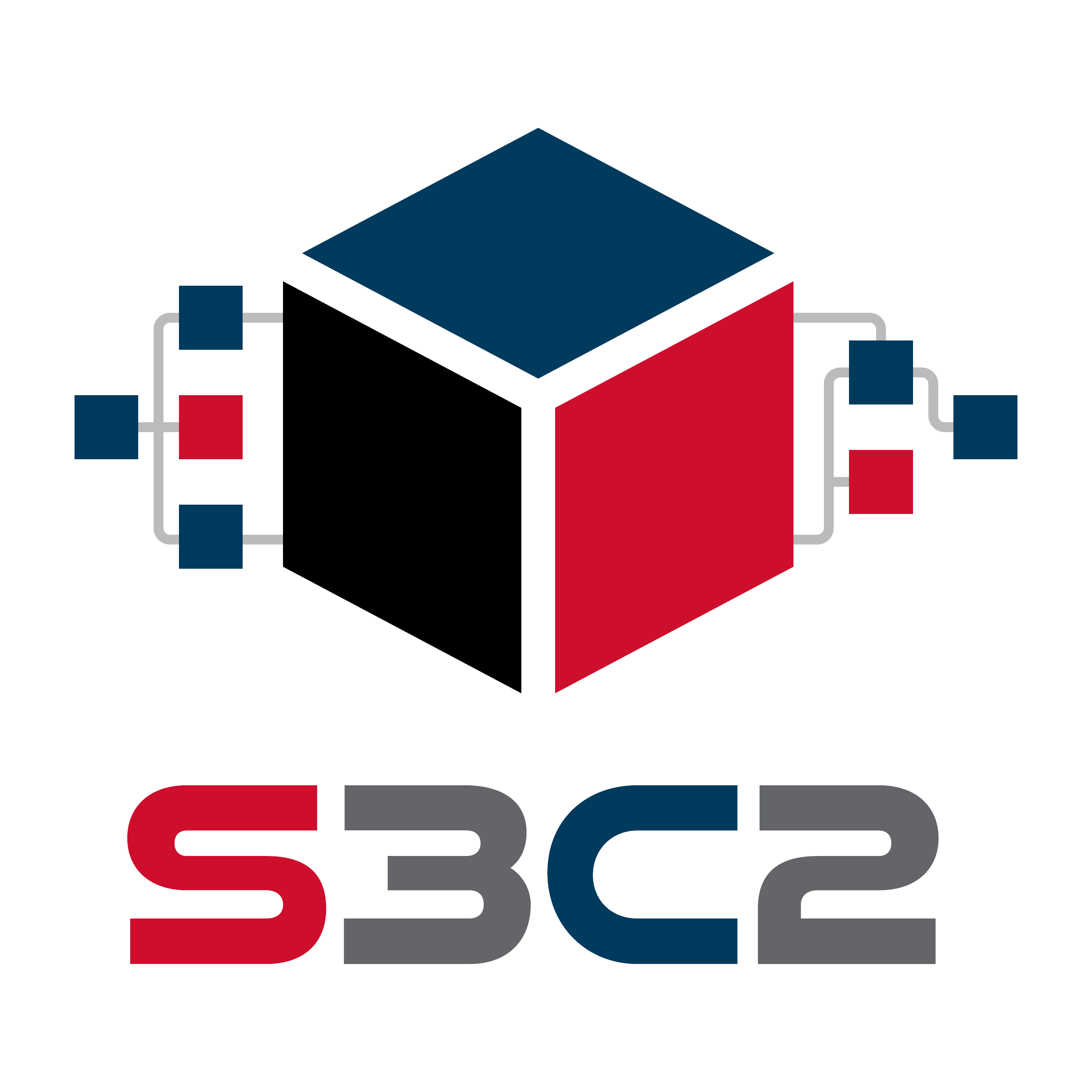}}; 
\end{tikzpicture}

\section{Overview}

Recent years have shown an increase in cyber attacks targeting less secure elements in the software supply chain and causing fatal damage to businesses and organizations. Past well-known examples of software supply chain attacks are the SolarWinds or log4j incidents that have affected thousands of customers and businesses. The US government and industry are both equally interested in enhancing software supply chain security. On September 30, 2022, three researchers from the NSF-supported Secure Software Supply Chain Center (S3C2)\footnote{\url{https://s3c2.org}} conducted a one-day Secure Software Supply Chain Summit with a diverse set of 19 practitioners from 17 companies.  The goal of the Summit is to enable sharing between industry practitioners having practical experiences and challenges with software supply chain security; to help form new collaborations between industrial organizations and researchers; and to identify research opportunities.

Summit participants were recruited from 17 companies, intentionally in diverse domains and having various company maturity levels and sizes \textemdash 12 major corporations, 3 medium size companies, 1 start up, and 1 software foundation representative).  Except for the host company with three participants, all companies could only have one participant.  Attendance is limited to one per company to keep the event small enough that honest communication between participants can flow. The Summit was conducted under the Chatham House Rules, which state that all participants are free to use the information discussed, but neither the identity nor the affiliation of the speaker(s), nor any other participant may be revealed. As such, none of the participating companies are identified in this paper.  

  The Summit consisted of one keynote presentation and six panels. Prior to the Summit, participants completed a survey to vote on the topics of the six panels.  As such, the panel topics represent the challenges faced by practitioners. Based upon personal preferences expressed in the survey, four participants were selected to begin each 45-minute panel discussion with a 3-5 minute statement.  The remaining minutes of each panel were spent openly discussing the topic.  The questions posed to the panelists appear in the Appendix.        

  The three researchers (two professors and one PhD student) and several participants took notes on the discussions.  The PhD student created a first draft summary of the discussion based on these notes.  The draft was first reviewed by the two professors at the Summit and then by the four other authors of this paper, who are also S3C2 researchers and experts in software supply chain security.

  The next seven sections provide a summary of the Secure Supply Chain Summit.

\section{Executive Order}
Executive Order (EO) 14028 issued on May 12, 2021 \cite{EO}, charges organizations supplying critical software to the US government to improve the security and integrity of their software and the software supply chain.  Most organizations need to make procedural, operational, and cultural changes as a result.


\subsection{Changes}
Some of the participants had met with NIST, CISA, and the administration to shape both the Executive Order draft itself and also to get support and clarity on fulfilling the EO requirements. Many internal security initiatives and existing practices already satisfy and match these requirements. Some practitioners stated that the EO has only been a catalyst for a security focus they wanted to have anyway. 

\subsection{Challenges}
The Summit participants found the Executive Order was ambiguous and not very actionable. They also felt that EO follow-on memos and documents did not lead to more clarity. 
Some practitioners also have concerns about achieving compliance at large. The EO also exposes smaller companies, non-profits, and open-source projects to compliance. These organizations typically do not have the resources to hire whole departments of specialists who understand how to audit against government requirements or to hire big IT shops who can provide a common platform to their developers to fulfill the requirements. Self-attestation might not be a feasible alternative for some companies as there might be a shortage of personnel to validate compliance.
Practitioners also believe that achieving compliance may require developers to make many procedural and cultural changes. Developers have come to expect some freedom in their development processes, particularly in large companies that acquire other companies. Complying with the EO and the self-attestation requirements may require a software development process that complies with the NIST Special Publication 800-218 Secure Software Development Framework (SSDF) \cite{SSDF} or the Supply-chain Levels for Software Artifacts (SLSA) \cite{SLSA} framework, which may take away some of these freedoms. 

\subsection{Open Questions}
At the end of the panel, one open question remained:

\begin{itemize}
    \item What are the consequences of non-compliance? How likely is non-compliance? Is Executive Order 14028 just something to create legal issues for companies that are not compliant?

\end{itemize}


\section{Software Bill of Materials (SBOM)}
An SBOM is a nested inventory of ``ingredients'' that make up the software component or product that helps to identify and keep track of third-party components of a software system. The EO states that any company that sells software to the federal government is mandated to issue a complete SBOM that complies with the National Telecommunications and Information Administration (NTIA) Minimal Elements \cite{NtiA}.  


\subsection{Benefits for Producers and Consumers}
The main discussions of this panel centered around the benefits that SBOMs could provide for producers and consumers. Most industry practitioners benefit from inventory disambiguation and believe that SBOMs can help establish customer trust by providing increased transparency and integrity of their deliverables. Customers can look for extraneous content, identify unwanted third-party packages, and enable local reasoning about vulnerabilities through SBOMs. Consumers can pinpoint affected components and assess the risk associated with the vulnerability. Practitioners further believe that SBOMs could lead to a heightened watch for a corresponding patch/VEX/VDR from the vendor and can be used to hold the involved persons accountable by enforcing legal or license compliance. An SBOM can be accompanied by a Vulnerability Exploitability eXchange (VEX) \cite{VEX} addendum which is a form of a security advisory that indicates whether a
product or products are affected by a known vulnerability or vulnerabilities. A Vulnerability Disclosure Report (VDR) is an attestation of all vulnerabilities affecting a product, or its’ dependencies, along with an analysis of the impact. According to the Summit participants, SBOMs mostly help a software producer with the questions ``am I affected?'' or ``what is affected?'' when a vulnerability is discovered, such as log4j.

\subsection{Challenges}
Participants have various concerns regarding the usefulness and maturity of SBOMs. One of the biggest hurdles of SBOMs is the overall immaturity of tooling relative to SBOM consumption. The maturity of producing is still more advanced than the maturity of consuming SBOMs, especially in an automated way. SBOM-related features (e.g., consuming, outputting additions, signing, transferring, and validating an SBOM in an automated way) should preferably be added to existing tools. A current area of innovation is \href{https://omnibor.io/}{OmniBOR}, which can construct a compact, verifiable Artifact Dependency Graph. 

Practitioners further mentioned concerns about demonstrating an SBOM's soundness and correctness. Consistent and coherent information is crucial but very hard to receive. Consumers are still very reliant on third parties to provide a correct SBOM. However, first-generation SBOMs are still not trustworthy and require running regular scans. As of now, SBOMs are not giving any savings yet, creating an analysis of known or likely low-quality SBOMs. SBOMs can also be very large, too large for ingesting by some text editors.

Some practitioners also still doubt of the usefulness and contribution of SBOMs in addition to VEX, especially for already-security-conscious and active vendors as SBOMs do not provide a lot of contextual data. A simple SBOM manifesting component names and versions does not necessarily cover interactions, impact, usage, exposure, or why a seemingly better component version is not in use.

However, most participants believe that the EO does represent a good forcing function for producers to manage better the content they deliver. However, regulators have not established requirements correctly, and regulations must evolve. Following the ``SBOM lifestyle'' is a good starting point and can help build out the core capability to understand supply chains in the future.

\begin{quote}
    \textit{``In 5 or 10 years of SBOM lifestyle we are going to be in better shape, living an SBOM lifestyle.''}\\
        -- Summit attendee.
\end{quote}

\subsection{Open Questions}
At the end of the panel, some open questions remained:

\begin{itemize}
    \item Should package managers be updated to emit standard format SBOM?
        \item What will the government do with all the SBOMs they are requiring?  Consuming SBOM is not required and is currently missing in the value chain.  
        Do SBOMs need to be signed?  
        Will there be a public repo for SBOM?
    \item Is it okay or a problem for an SBOM to potentially over-report (e.g., include non-shipped build dependencies?) or under-report (e.g., contractually may not be permitted to disclose usage of an internal component)?
\end{itemize}

\section{Updating Vulnerable Dependencies}

Most software uses a plethora of third-party dependencies that provide common functionality and help developers with productivity. However, these dependencies adds an additional layer of complexity and leads to an ecosystem of (transitive) dependencies that each software replies on. A security vulnerability in a third-party dependency can lead to cascading issues and needs to be updated with the newly released version fix as soon as possible. Companies commonly rely on different strategies and tools when updating vulnerable dependencies. 


\subsection{Processes and Challenges}
Practitioners mentioned scanning for vulnerabilities by using Software Composition Analysis (SCA) tools.~
The scaling factor of transitive dependencies causes dependency graphs to be large and ever-changing. Multiple practitioners mentioned feeling overwhelmed by the number of vulnerabilities identified by the tools in direct but also transitive dependencies. After updating all vulnerabilities, a tool run the next day will may identify more new vulnerabilities. This makes it extremely hard for practitioners to catch up and forces them to triage and prioritize  certain vulnerabilities based upon severity, CVSS score, or cost-benefit analysis. 
Tooling to automatically patch dependencies (e.g., Dependabot’s automated pull requests) is available. However, auto-patching is often not accepted and may not be very realistic for some companies as they cause pull request (PR) flooding and still require humans to review all patch updates manually. Defining the appropriate version is also not always straightforward and needs to be done on the basis of complex criteria as the newest version may be too close to the leading edge and some companies prefer to wait to see if new problems surface with the update. 

Other challenges arise when deploying an update. Version changes may need to be done multiple times because there is not one source of truth for dependency usage as different SBOM or SCA tools give different results. Updating the dependency in a repository also does not mean the change is universally consumed by the continuous integration (CI) and fully deployed to all runtime locations. Vulnerable components may also need to be updated and fixed multiple times, as occurred with past incidents (e.g., log4j).  Developers are oftentimes uncertain about the next steps and do not feel responsible. Clear ownership and actionable advice are direly needed (e.g., ``You must move project N from version A to B.'').

Practitioners believe that understanding transitive dependencies could help in improving human operation by solving multiple vulnerabilities by updating one higher-level dependency. One practitioner suggested grouping and updating vulnerable dependencies weekly to give their direct dependencies a chance to update the transitive dependencies. 

Dismissing updates of a known vulnerable version due to criteria showing the vulnerable functionality is not in use may hide the problem of potential future usage of the functionality. Some customers are requiring all vulnerabilities to be fixed, regardless of usage, reducing the goal of having a VEX document.

\subsection{Open Questions}
At the end of the panel, some open questions remained:

\begin{itemize}
    \item Is dependency version pinning good or not?
    \item Should a different level of review be applied to dependency updates? For example, can a Dependabot PR be considered one ``reviewer?''
    \item Must a dependency be updated? Differentiating ``update available'' and ``update must be rolled out'' may depend on subjective or at least relative / non-universal criteria that might reduce toil. 
    \item Is there an aging metric (``how old are the dependencies,'' ``how old are my deployments'') which could be leveraged to ease planning or encourage the practice of staying more fresh.  One metric discussed was ``vuln agility'' whereby if the SLA said you needed to fix a vulnerability within 7 days and you do it in 1 day, your score is 1/7. Lower numbers for vuln agility are better. 
    \item Who is the owner of a dependency and its lifecycle management? Mono Repositories may make it easier to find code and usages, but diffuse ownership and responsibility. Who has the power to declare a vulnerability as WILLNOTFIX? Where are exemptions recorded and re-evaluated?
    \item Are vulnerabilities in transitive dependencies many layers deep less risky than vulnerabilities in a direct dependency? A general consensus using log4j as an example is that it did not matter where the dependency was in the dependency graph, it needed to be fixed.
    \item How often are there build breaks if auto-updates are done?
\end{itemize}

\section{Detecting Malicious Commits}
Actors of past software supply chain attacks (e.g., SushiSwap) use malicious commits to submit unauthorized changes to the source repository. Detecting and discerning these malicious commits is not always straightforward as attackers often use obfuscated code, steal authentication credentials, or use impersonation strategies to deceive and put malicious code changes through the system. 


\subsection{Signals}
Multiple practitioners believe that taking a closer look at the committer’s behavior might be helpful in discerning malicious behavior. For instance, a committer’s activity, reputation as well as uncommon behavior (e.g., big vs small changes, critical vs ancillary fixes) can signal suspicious or malicious commits. Machine Learning (ML) could be applied to create appropriate tooling. However, analyzing and interpreting committers’ activity might be hard for certain valid scenario groups (e.g., anonymous or new contributors).

\subsection{Challenges}
Several practitioners agree that it is hard to define malicious commits as the committer's intention is not always clear. Innocent coding mistakes can hardly be distinguished from maliciously intended ones and maliciously intended ones can be covered and concealed as innocent coding mistakes. Malicious commits could be obfuscated code, bugs, signed commits with stolen credentials or compromised identity, or even unsigned commits. New security frameworks such as SLSA require two-party reviews at higher levels. However, actively obfuscated code or code that follows certain patterns (e.g., typosquatting, backdoors, dependency confusion) is likely to get past human code review. Some of them have been spotted and remediated automatically with \href{https://github.com/microsoft/OSSGadget}{OSSGadget}. Some practitioners believe that general operational patterns, threats, and mitigations need to be considered beyond just the code. Using signed commits and strong identity checks (e.g., 2FA) on platforms can help but cannot be used as sole protection against malicious commits since malicious actors can replace authentication mechanisms and secrets after compromising the human or one of their devices. Moreover, companies need to consider the developer's time and productivity. Requesting re-authentication for every code commit is very time-consuming and unacceptable. 

\subsection{Open Questions}
At the end of the panel, some open questions remained:

\begin{itemize}
    \item Sandboxing pipelines vs sandboxing inbound source code commits: Should pre/post-install scripts be disabled? Or disable all code install ability? On what machines?
    \item What if it is a critical ``everything is down'' bug at 3 am or one of your founding engineers’ identity was stolen? Odds are the two +1 reviews come through rapidly (or are skipped explicitly?). Reviews cannot be the saving grace.
\end{itemize}

\section{Secure Build and Deploy}
Various build platforms and CI/CD tools support developers in automating the parts of software development related to building, testing, and deploying. These platforms further help in enhancing software build integrity by establishing documented and consistent build environments, isolating build processes, and generating verifiable provenance. 


\subsection{Secure Build and Deploy Processes}
Most practitioners feel comfortable in securing the deployment process and are more worried about the build process. Almost all practitioners have logging and monitoring in place to provide more transparency. One practitioner is using a third-party auditor for monitoring. Others issue credentials and use those for publishing artifacts and testing. They generally have a chain of custody with many checks and signing. Metadata is always collected in the backend system and analyzed in CI. There still seem to be challenges in managing endpoint security since developers oftentimes have several devices. Hardening the build system might also cause additional costs for developers since they have to adjust to new practices.

Participants believe that the build cluster should be segregated from the production cluster so developers cannot access it. There should be a better separation between CI secrets and CD secrets and individuals should be given fewer rights to have more assurance that CD secrets are not shared back. 

Overall, companies seek guidance on secure build and deployment from the security framework SLSA, though most companies may still operate on lower security levels. The 2021 Summits indicated that companies did not know how to secure build and deploy. In contrast, this group seems to be well aware of the next steps but indicated needing more time for implementation and reaching higher security levels.

\subsection{Reproducible Builds}
Reproducible Builds is a process of compiling software that ensures the resulting binary code can be reproduced bit-by-bit. The v0.1 version of the SLSA framework \cite{SLSA} included Reproducible Builds as a build requirement on higher levels, and several open-source software already provide or plan to provide reproducible builds. However, most practitioners at the summit believe that reproducible builds may not be very feasible to check whether a build has been tampered with as there still seem to be many challenges and concerns. As of now, only 20\% of builds match bit-to-bit. Most practitioners would like to utilize Reproducible Builds but find it hard to implement as certain programming languages do not seem practical and introduce randomness or non-deterministic data (e.g., Docker images with timestamps). Other associated aspects such as computation and validation time and sustainability (environmental cost of building multiple times) also need to be considered.

Currently, no other established alternatives to Reproducible Builds have been identified. Some practitioners at the Summit suggested checking for functional reproducibility between versions to see behavior changes, switching to a safer language, or working on publishing hashes only.

\subsection{Open Questions}
At the end of the panel, some open questions remained:

\begin{itemize}
    \item What are some other alternatives to reproducible builds to achieve the goal of identifying tampering in the build process?
\end{itemize}

\section{Supply Chain Standards, Guidelines, Frameworks}
Software security standards, guidelines, and frameworks (e.g., SSDF \cite{SSDF}, NIST 800-161 \cite{800_161}, OWASP Software Component Verification Standard (SCVS) \cite{SCVS}, Supply-chain Levels for Software Artifacts (SLSA) \cite{SLSA}, and Software Supply Chain Consumption Framework (S2C2F) \cite{S2C2F}) have emerged to guide what organizations can do to reduce software supply chain risk. Participants discussed the extent to which they were guided or complied with these standards, guidelines, and frameworks.   


\subsection{Followed Standards}
Most participants started working towards fulfilling higher level numbers in the SLSA \cite{SLSA} security framework. A cross-industry collaboration has recently released SLSA, a security framework for improving the end-to-end integrity of a software artifact throughout its development lifecycle, particularly focused on the build process. SLSA consists of different level numbers and a checklist of standards and controls to prevent tampering, improve integrity, and secure packages and infrastructure. Participants believed that SLSA focuses on the immediate need of the open source supply chain and can pose as a great starting point for incrementally reaching general compliance with many of the government-required standards.  Challenges lie in explaining certain details and source requirements of SLSA to developers.  Participants also desired SLSA tool support. A collaboration with GitHub has led to new sets of tools and features like native builders and verifiers which give easy path to generating and verifying SLSA provenance files. 

Other participants followed the Cloud Native Development Foundation (CNCF) Software Supply Chain Best Practices \cite{CNCF}, NIST SSDF \cite{SSDF}, and OWASP SCVS \cite{SCVS}. The OpenSSF-endorsed S2C2F \cite{S2C2F} outlines and defines how to securely consume OSS dependencies, such as NuGet and npm, into the developer's workflow.

\subsection{Challenges}
Some practitioners still seem to have a hard time deciding on a standard as there are several disparate sources of standards and guidelines. 
Participants also doubted in the cryptographical viability of those standards. One participant noted that there also seem to be missing standards for certain aspects, such as sandboxing, used languages, or attributes of languages. Letting the government define these standards is hard to accomplish and risks them getting it wrong as they may not have adequate expertise and experience.

Practitioners further expect to face challenges when demonstrating compliance to a given standard (e.g., SBOM \cite{NtiA}) as it is still unclear how compliance will be validated and which specific forms will be requested to demonstrate compliance. 

Overall, most participants believe that standards can help in providing companies with a clear roadmap and guidance. However, they should always be accompanied by a level number to enable companies to define measurable goals and track outcomes.

\section{Misc discussion}
The keynote raised awareness of many Google and industry efforts to secure the software supply chain. There has been an influx of new projects, frameworks, and tools to strengthen the software supply chain security.

Google has created a dedicated Open Source Security Team (GOSST) and further announced its Assured Open Source Service (Assured OSS).\footnote{\url{https://cloud.google.com/assured-open-source-software}} With this service, customers can get components that have gone through Google's OSS code testing and vulnerability management processes. GOSST was created in 2020 to work with the open-source community at large to make the open-source software that everyone relies on more secure. GOSST is involved in many initiatives, such as Sigstore\footnote{\url{https://www.sigstore.dev/}} and OpenSSF. The Sigstore initiative provides a new way of signing software artifacts such as release files, container images, binaries, and bill of material manifests without requiring key management for users. Sigstore uses open-source technologies, like Fulcio, Cosign and Rekor, and handles digital signing, verification and checks for provenance to make it safer to distribute and use open-source software. Sigstore is hoped to have a similar industry impact to ``let’s encrypt.''\footnote{\url{https://letsencrypt.org/}} 

The OpenSSF initiative has further launched AllStar,\footnote{\url{https://github.com/ossf/allstar}} a project that provides automated continuous enforcement of security best practices for GitHub projects and uses continuous monitoring of security health metrics. Allstar is a companion of the OpenSSF Scorecard project\footnote{\url{https://github.com/ossf/scorecard}} but can still profit from better integration to the scorecard project. New tools that can aid in evaluate fuzz testing tools properly and rigorously, such as OSSFuzz, ClusterFuzzLite, or FuzzBench. 

\section{Executive Summary}
Some practitioners stated that the
EO \cite{EO} has only been a catalyst for a security focus in their organizations.  However, these participants also were concerned with compliance and participarly with the effect of the EO on smaller organizations and on the resulting procedural and cultural changes.  Overall, participants were supportive of the benefits of producing and consuming SBOMs \cite {NtiA}, but noted that tools to consume SBOMs are lagging behing those to produce an SBOM.  Discussions about processes for updating vulnerable direct and transitive dependencies revealed that most organizations are overwhelmed by the number of vulnerabilities identified by SCA tools with no participant indicating a recommended process to share.

Attackers are increasingly injecting malicious components into ecosystems which are used to launch attacks.  Participants commiserated on the difficulties in detecting malicious commits as the committer’s intention is not always clear. Innocent coding mistakes can hardly be distinguished from maliciously injected vulnerabilities. Organizations are adopting practices of the SLSA \cite{SLSA} framwork to reduce supply chain attackes through the build infrastructure. Discussions of the use of reproducible builds for the detection of tampering through the build system generally revealed challenging implementing this practice in mature organizations.

\begin{acks}
A big thank you to all Summit participants. We are very grateful for being able to hear about your valuable experiences and suggestions. The Summit was organized by Laurie Williams and Yasemin Acar and was recorded by Mindy Tran.  This material is based upon work supported by the National Science Foundation Grant Nos. 2207008, 2206859, 2206865, and 2206921.
These grants support the Secure Software Supply Chain Summit (S3C2), consisting of researchers at North Carolina State University, Carnegie Mellon University, University of Maryland, and George Washington University. Any opinions expressed in this material are those of the author(s) and do not necessarily reflect the views of the National Science Foundation.
\end{acks}

\bibliographystyle{ACM-Reference-Format}
\bibliography{literature}

\appendix

\section{Full Survey Questions for Panel}
\label{questions}
\begin{enumerate}
    \item What will/can SBOMs actually achieve? Are they a waste of time?  How can they be leveraged/used?
    \item What is your process for updating vulnerable dependencies Do you always keep up-to-date?  What kind of testing or other strategies do you use  before updating to a new version?  How is the output of an SCA tool used?
    \item How can malicious commits be detected? What do you think signals a suspicious/malicious commit?  What role does the ecosystem play in detecting malicious commits?
    \item What is being done (or should be being done) to secure the build and deploy process/tooling pipeline?  Are you working toward reproducible builds?
    \item What changes is your company making in relation to the Executive Order?
    \item What standards do you follow and/or use for guidance for secure software supply chain?  Examples:  SSDF, NIST 800-161, SLSA.
\end{enumerate}

\end{document}